\documentclass{pasj00}
\draft

\begin{document}
\SetRunningHead{J. K. Ishitsuka et al. PASJ 53-6-Water Masers around R
 Crateris}{Running Head}
\Received{2000/12/08}
\Accepted{2001/09/24}

\title{VLBI Monitoring Observations of Water Masers Around the Semi-Regular
Variable Star \mbox{R Crateris}}

 \author{%
   Jos\'e K. \textsc{Ishitsuka },\altaffilmark{1}
   Hiroshi \textsc{Imai},\altaffilmark{2,3}
   Toshihiro \textsc{Omodaka},\altaffilmark{4}
   Munetaka \textsc{Ueno},\altaffilmark{1} \\
   Osamu \textsc{Kameya},\altaffilmark{2,3}
   Tetsuo \textsc{Sasao},\altaffilmark{3}
   Masaki \textsc{Morimoto},\altaffilmark{5} \\
   Takeshi \textsc{Miyaji},\altaffilmark{3,6}
   Jun-ichi \textsc{Nakajima},\altaffilmark{7}
   and
   Teruhiko \textsc{Watanabe}\altaffilmark{4}} 
 \altaffiltext{1}{Department of Earth Science and Astronomy, University of
 Tokyo, Komaba, Tokyo 153-8902}
 \email{pepe@chianti.c.u-tokyo.ac.jp}
 \altaffiltext{2}{Mizusawa Astrogeodynamics Observatory, National
 Astronomical Observatory, Mizusawa, Iwate 023-0861}
 \altaffiltext{3}{VERA Project Office, National Astronomical Observatory,
 Mitaka, Tokyo 181-8588}
 \altaffiltext{4}{Faculty of Science, Kagoshima University, Kagoshima,
 Kagoshima 890-0065}
 \altaffiltext{5}{Nishi-Harima Astronomical Observatory, Sayo, Hyogo
 679-5313}
 \altaffiltext{6}{Nobeyama Radio Observatory, National Astronomical
 Observatory, Minamisaku, Nagano 384-1305}
 \altaffiltext{7}{Kashima Space Research Center, Communications Research
 Laboratory, Kashima, Ibaraki 314-0012}


\maketitle

\begin{abstract}

We monitored water-vapor masers around the semi-regular variable star
\mbox{R Crateris} with the Japanese VLBI Network (J-Net) at the \mbox{22
GHz} band during four epochs with intervals of one month.  The relative
proper motions and Doppler-velocity drifts of twelve maser features were
measured.  Most of them existed for longer than \mbox{80 days}.  The 3-D
kinematics of the features indicates a bipolar expanding flow.  The
major axis of the asymmetric flow was estimated to be at P.A.  =
136$^{\circ}$.  The existence of a bipolar outflow suggests that a Mira
variable star had already formed a bipolar outflow.  The water masers
are in a region of apparent minimum radii of 1.3 $\times$ 10$^{12}$ m
and maximum radii of 2.6 $\times$ 10$^{12}$ m, between which the
expansion velocity ranges from 4.3 to 7.4 km s$^{-1}$.  These values
suggest that the water masers are radially accelerated, but still
gravitationally bound, in the water-maser region.  The most positive and
negative velocity-drifting features were found relatively close to the
systemic velocity of the star.  We found that the blue-shifted features
are apparently accelerated and the red-shifted apparently decelerated.
The acceleration of only the blue-shifted features seems to be
consistent with that of the expanding flow from the star.

\end{abstract}

\section{Introduction}

The gas dynamics of circumstellar envelopes (CSEs) during the latest
phase of the asymptotic giant branch (AGB) stars is one matter of debate
to elucidate the physical mechanisms of energetic mass loss of evolved
stars.  The AGB stars evolve in a sequence of Mira variables, IRC (or
AFGL) objects, OH/IR stars, proto-planetary nebulae (PPNs), and
planetary nebulae (PNs) (\cite{Taka}).  Recent HST infrared images (e.g.
\cite{Sahai}) have shown asymmetric or bipolar outflows around the
central object of PNs; PNs show axisymmetric morphologies in over 50\%
of the cases (e.g.  \cite{Sahai}; \cite{Manc}).  On the other hand, the
CSEs of AGB stars are generally spherically symmetric (\cite{Manc}).
The most intriguing question is when the change of morphology occurs
between the AGB and PN phases.

Water-vapor masers have been well observed around many evolved stars.  A
cluster of water-maser features exists at 10$^{12}$ $-$ 10$^{13}$ m from
the central star (\cite{Joh}; \cite{Lane-a}).  Maser action is predicted
in the inner part of the CSE, and the size of the maser region increases
with the stellar mass-loss rate (\cite{Cook}).  Distributions of water
masers in CSEs are complex, and change with time (e.g.  \cite{Lane};
\cite{Joh}).  The time variation was reported on a scale ranging from a
few weeks to a few years (\cite{Bo-a}).  The VLA, VLBA, and MERLIN
observations have monitored the water-maser features:  the epochs were
widely scattered from a half year to several years (\cite{Mar}).
Measurements of the relative proper motion of compact maser features
with high spatial resolution can be used to investigate the structure of
CSEs and to trace those asymmetries or bipolarities (\cite{Mar}).
Measurements of 3-D velocities of water-maser features as well as their
radial-velocity drifts (\cite{Im-a}) enable us to elucidate the dynamics
of the mass-loss process in more detail.  For this purpose, monitoring
observations within intervals of a few weeks are necessary to verify
whether the same maser features are observed within the epochs.

R Crt is a semi-regular variable star of SRb type in spectral
classification M7 (\cite{Khol}).  R Crt is classified as a Mira variable
from a color-color diagram of $IRAS$ observations (\cite{Taka}).  The
$V$-band magnitude varies between 9.8 and 11.2 mag with a pulsation
period of 160 days.  $^{12}$CO($J$=2$-$1) and ($J$=1$-$0) and
$^{13}$CO($J$=2$-$1)(\cite{Kaha-a}) as well as OH and SiO masers
(\cite{Le-S}; \cite{Jew}) were detected at the envelope.  The mass-loss
rate of R Crt was estimated to be 1.0 $\times$ 10$^{-6}$ $M$$_\odot$
yr$^{-1}$ (\cite{Kaha-a}) from  CO observation.  The systemic velocity
for \mbox{R Crt} is 10.8 km s$^{-1}$ (\cite{Bo-b}), which was estimated
from the center velocity of the velocity coverage of a CO observation
profile.  Water masers around R Crt were first detected by \citet{Dick}.
In this paper we present the results of monitoring observations for
water masers around R Crt.  We compared the positions of the water-maser
features among epochs separated by a few weeks, and measured radial
velocities throughout the epochs.  Based on the 3-D motions and
radial-velocity drifts, we discuss the dynamics of the mass-loss flow
from this star.

\section{Observations and Data Reduction}

Monitoring observations of \mbox{R Crt} at R.A.  (B1950.0) =
\timeform{10h58m6.14s} and Decl.\ (B1950.0) =
${-}$\timeform{18D03'18.27"} were made by using the \mbox{Japanese VLBI}
Network (J-Net, \cite{Omo}) from 1998 March to June, at four epochs with
an interval of a few weeks.  J-Net is comprised of three of the National
Astronomical Observatory of Japan's telescopes, including the
\mbox{45-m} telescope at Nobeyama Radio Observatory \footnotemark[(*)],
the \mbox{10-m} telescope at Mizusawa, and the \mbox{6-m} telescope at
Kagoshima, as well as one of the Communications Research Laboratory's
telescopes, the \mbox{34-m} telescope at Kashima.  \mbox{Table
\ref{tab:obs-station}} gives the characteristics of the telescopes, and
table \ref{tab:obs-epoch} the observations.

\footnotetext[(*)]{Nobeyama Radio Observatory is a branch of the National
      Astronomical Observatory, operated by the Ministry of
      Education, Culture, Sports, Science, and Technology(, Japan).}

\begin{table*}
\begin{center}
\caption{J-Net telescopes that participated in the observations.}
\label{tab:obs-station}
\vspace{4pt}
{\scriptsize
\vspace{6pt}
\begin{tabular*}{13.0cm}{@{\hspace{\tabcolsep}
\extracolsep{0pt}}p{6pc}ccccc}
\hline\hline\\[-6pt]
Station & Designation & Diameter (m)  & Ap. efficiency (\%) & $T_{sys}$ 
at zenith (K) \\
[4pt]\hline\\[-6pt]
Mizusawa \dotfill & M & 10 & 36 & 140 -- 330 \\
Kashima \dotfill & O & 34 & 57 & 280 -- 340 \\
Nobeyama \dotfill & N & 45 & 63 & 240 -- 290 \\
Kagoshima \dotfill & K & 6 & 40 & 170 -- 220 \\
\\[-3pt]
\hline\\[-6pt]
\end{tabular*}

}
\end{center}
\end{table*}

In each epoch \mbox{R Crt} was observed at 5 to 8 intervals for 20 to 40
minutes alternating a scan for 5 to 10 minutes toward \mbox{3C 273B} as
a calibrator.  The observed signals were recorded with the VSOP terminal
(\cite{Kawa}) in one base band channel with a band width of \mbox{16
MHz} in \mbox{2 bits} per sample.  The recorded data were correlated
using the \mbox{Mitaka FX Correlator} (\cite{Chik}).  The output of the
correlator had an \mbox{8 MHz} bandwidth with \mbox{512 spectral
channels}, corresponding to a Doppler velocity resolution of 0.105 km
s$^{-1}$.  The velocity coverage was from $V_{\rm{LSR}}$ = \mbox{${-}$16
km s$^{-1}$} to \mbox{36 km s$^{-1}$} centered on the water-maser line.

We used the NRAO's AIPS software to calibrate, image and fit the
brightness distribution.  The strongest water maser component detected
at $V_{\rm{LSR}}$ = 15.03 km s$^{-1}$ was used as phase reference in
each set of observations.  The synthesized beam of the array was 8 mas
$\times$ 3 mas with a position angle of ${-}$ 40$^{\circ}$ in natural
weight.  The typical rms thermal noise level in the channel maps ranged
from 34 mJy to 80 mJy between the first and last epochs (see table
\ref{tab:obs-epoch}).  The accuracy of the intensity scale is estimated
to be better than 50$\%$.  The accuracy of the relative angular position
ranged from 0.2 mas to 0.9 mas in the R.A.  direction and 0.2 mas to 1.1
mas in the declination direction, in a single velocity channel.


\begin{table*}
\begin{center}
\caption{Time table of the monitoring observation.}
\label{tab:obs-epoch}
\vspace{4pt}
{\scriptsize
\vspace{6pt}
\begin{tabular}{cccccc}
\hline\hline\\[-6pt] \vspace{1pt} 
Experiment & Epoch & Stations & Observation & Typical rms thermal &
Number of detected \\
name & (1998) &  & duration (hr) &
noise (mJy) &  features  \\
[4pt]\hline\\[-6pt]
j98090 \dotfill & March 31 & MKNO & 10 & 34 & 14 \\
j98116 \dotfill & April 26 & MKNO & 10 & 55 & 13 \\
j98148 \dotfill & May 28 & MKNO & 9 & 70 & 12 \\
j98170 \dotfill  & June 19 & MKNO & 8 & 80 & 5 \\
\\[-3pt]
\hline\\[-6pt]
\end{tabular}
}
\end{center} 
\end{table*}

A maser spot is a single-velocity component, while a maser feature is a
group of maser spots within several mas in positions and 0.1 $-$ 0.5 km
s$^{-1}$ in velocities.  The definitions of `maser spot' and `feature'
used in this paper are the same as those used in \citet{Im-b} and
references therein.  These definitions are the same as those given by
\citet{Gwin}.  The positions of the brightness peaks in the maser spots
were estimated using the AIPS task JMFIT.  We traced the position and
velocity at the brightness peak in each of the features.  We determined
the velocity at the brightness peak of a feature by quadratic fitting of
the observed intensities against the velocities of the strongest three
maser spots in the feature.  The position of the feature at the
calculated peak velocity was estimated on the line-connecting positions
of two velocity-adjacent spots assuming a constant velocity gradient
between the two spots.

\section{Results and Discussion}

\begin{figure}
  \begin{center}
    \FigureFile(150mm,180mm){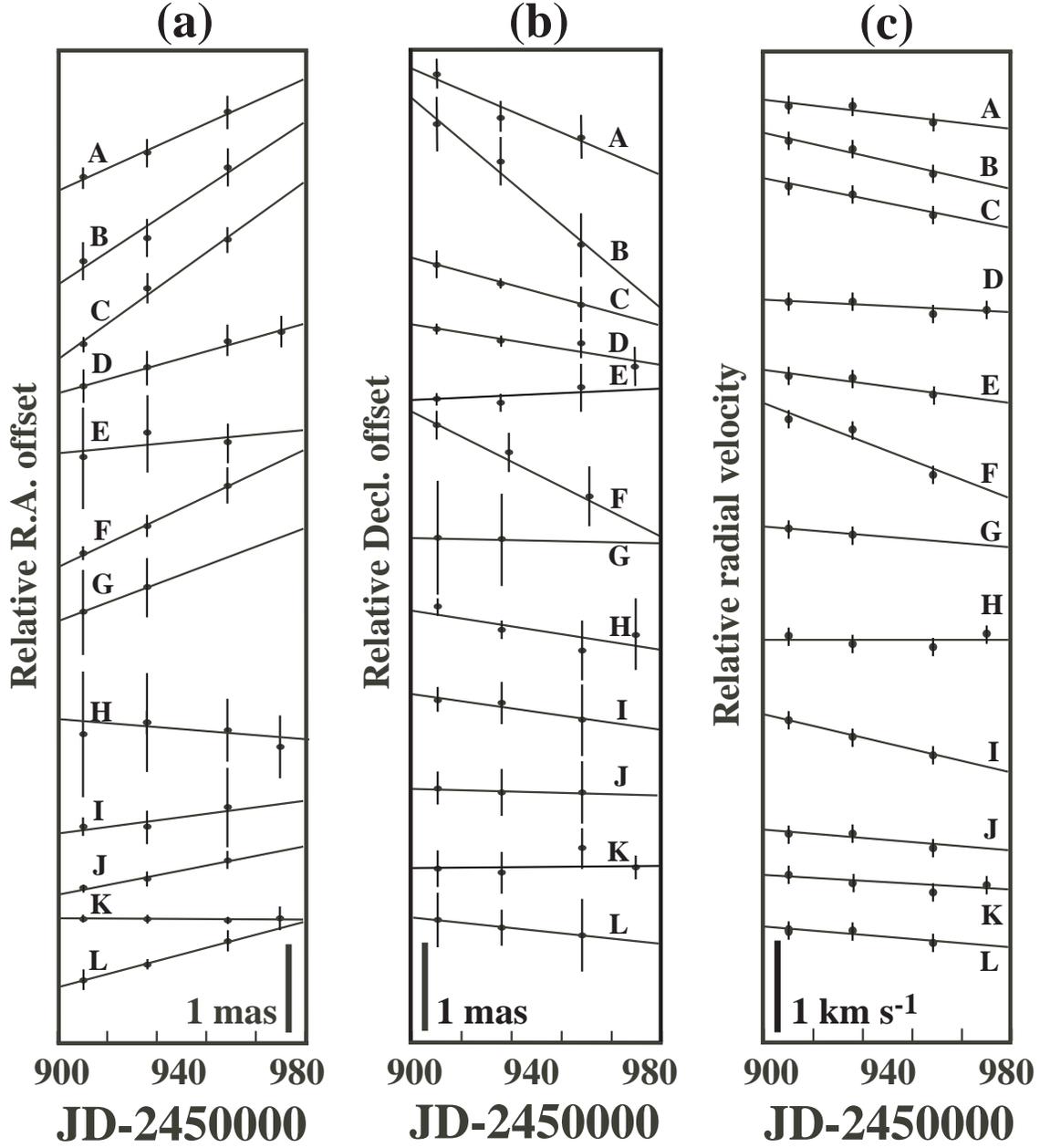}
  \end{center}
  \caption{Time variation of \mbox{R Crt} maser features in (a)
           \mbox{R.A.}, (b) Decl., and (c) the Doppler velocity.
           Features A $-$ G are blue-shifted features and H $-$ L 
           red-shifted features, with respect to the
           systemic velocity, $V_{\rm{LSR}}$ = 10.8 km s$^{-1}$.}
  \label{fig:P.M.}
\end{figure}

For proper-motion measurements, the four maps were superposed so as to
situate the position-reference feature ($V_{\rm{LSR}}$ = 15.03 km
s$^{-1}$) at the same position in the superposed single map.  Figure
\ref{fig:P.M.}  shows the observed proper motions of the maser features
in the R.A.  and Decl.  directions.  We also measured the
Doppler-velocity drifts of the maser features that are shown in figure
\ref{fig:P.M.}.  The uncertainty of the relative velocities of the maser
features was assumed to be 0.05 km s$^{-1}$, one half of our velocity
resolution.  Since the position-reference is also moving relatively, the
mean proper motion of the features was calculated and subtracted from
each of the proper motions.  \mbox{Table \ref{tab:water-masers}}
summarizes the relative positions, proper motions, Doppler velocities,
Doppler-velocity drifts, and peak intensities of the maser features.  A
letter is assigned to each of the features, and hereafter we use these
assignations.  We confirmed that the strongest features (D, H, and K)
were detected at four epochs, and existed for at least 80 days, while
others were detected during two or three epochs, possibly due to noisy
observation at the third and fourth epoch.

\begin{table*}
\caption{Obtained parameters of the water-maser features with proper motions.}
\label{tab:water-masers}
\vspace{4pt}
{\scriptsize
\begin{center}
\begin{tabular}{l@{ }r@{ }rr@{ }rr@{ }rr@{ }rr@{ }rr@{ }r@{ }r@{ }r} 
\hline\hline\\[-6pt] \vspace{1pt} 
& \multicolumn{2}{c}{Position} 
& \multicolumn{6}{c}{Relative velocity}
& \multicolumn{2}{c}{V$_{z}$ drift}
& \multicolumn{4}{c}{Peak intensity} \\ 
\hline \\[-6pt] \vspace{3pt}
Feature & $\Delta$X & $\Delta$Y & $\mu_{x}$ & $\sigma_{\mu_{X}}$ & $\mu_{y}$ 
 & $\sigma_{\mu_{Y}}$ & $V_{z}$\footnotemark[*] & $\Delta_{V_{Z}}$ 
& $\dot{V_{z}}$ & $\sigma_{\dot{V_{z}}}$ 
& Epoch 1 & Epoch 2 & Epoch 3 & Epoch 4 \\ \hline\\[-6pt] \vspace{5pt}
 & (mas) & (mas) & \multicolumn{2}{r}{(km s$^{-1}$)} 
 & \multicolumn{2}{r}{(km s$^{-1}$)}
 & \multicolumn{2}{r}{(km s$^{-1}$)}
 & \multicolumn{2}{r}{(km s$^{-1}$yr$^{-1}$)}
 & & \multicolumn{2}{r}{(Jy beam$^{-1}$)} \\ 
A ... \dotfill &     48.23 &    $-$68.58 &   2.55 &   2.35 &  $-$4.00 &   3.59
 &   $-$7.44 &   0.73 &   $-$1.37 &   0.45
  &      7.87 &      3.07 &        1.43 &     ...  \\
B ... \dotfill &     46.20 &    $-$64.82 &    0.84 &   7.45 &  $-$6.34 &  11.11
 &   $-$6.88 &   0.42 &   $-$1.27 &   0.99
  &       5.48 &       1.62 &     ...      &     ...   \\
C ... \dotfill &     65.52 &     $-$3.05 &   7.66 &   1.80 &   $-$2.38 &   3.06
 &   $-$5.41 &   1.09 &   $-$2.30 &   0.45
  &      2.09 &      6.32 &        1.92 &     ...  \\
D ... \dotfill &    $-$47.03 &     36.59 &  $-$0.25  &   2.19 &   0.93 &   1.36
 &   $-$1.81 &   0.76 &   $-$0.68 &   0.30
  &      1.45 &      5.61 &        2.32 &      2.31 \\
E ... \dotfill  &    $-$55.34 &     42.34 & $-$4.64 &   5.87 &   4.11 &   2.52
 &   $-$0.71 &   0.99 &   $-$1.55 &   0.45
  &      3.64 &       1.33 &        1.47 &     ...  \\
F ... \dotfill &     53.42 &    $-$37.67 &   2.84 &   2.07 &  $-$6.2 &   4.23
 &   $-$0.24 &   1.58 &   $-$4.32 &  0.45
  &      2.62 &      1.43 &        0.86 &     ... \\
G ... \dotfill &    $-$62.24 &     40.48 &   3.68 &  19.10 &    4.78 &  11.30
 &    0.42 &   0.36 &   $-$0.98 &   0.99
  &       2.18 &       0.73 &     ...      &     ... \\
H ... \dotfill &    $-$67.80 &     38.97 &   $-$7.96 &   7.02 &  0.33 &   1.30
 &    1.12 &   0.95 &    0.05 &   0.30
  &      8.93 &      3.91 &        2.18 &      2.97 \\ 
I ... \dotfill &    $-$72.40 &     38.78 &    0.67 &  13.00 &   $-$0.97 &   5.08
 &    1.91 &   0.73 &   $-$2.52 &  0.81
  &     2.19      &       1.41 &        0.64 &     ...   \\
J ... \dotfill &     20.09 &     46.81 &  $-$0.62 &   2.08 &   2.95 &   2.12
 &    2.72 &   0.81 &   $-$1.04 &   0.45
  &       1.76 &       1.29 &        1.16 &     ...  \\
K\footnotemark[\dag] ... \dotfill &      0.00 &      0.00 &  $-$5.54 &   0.59 &    4.46 &   0.95
 &    4.23 &   1.21 &  $-$0.73 &   0.30
  &      30.40 &      30.07 &        25.20 &     20.35 \\ \vspace{3pt}
L ... \dotfill &     $-$0.46 &     46.02 &    0.77 &   1.37 &   2.32 &   2.58
 &    6.56 &   1.09 &   $-$1.04 &   0.45
  &      8.98 &       2.17 &        0.90 &     ...    \\  
\hline \\[-6pt] 
\end{tabular}
\end{center}
\hspace*{1.0cm}
\footnotemark[*] Relative Doppler velocity with respect to the systemic 
velocity $V_{\rm{LSR}}$ = 10.80 km s$^{-1}$  \\
\hspace*{1.0cm}
\footnotemark[\dag] Position reference feature at $V_{\rm{LSR}}$ = 15.03
km s$^{-1}$  \\
}

\end{table*}

\begin{figure}
  \begin{center}
    \FigureFile(120mm,120mm){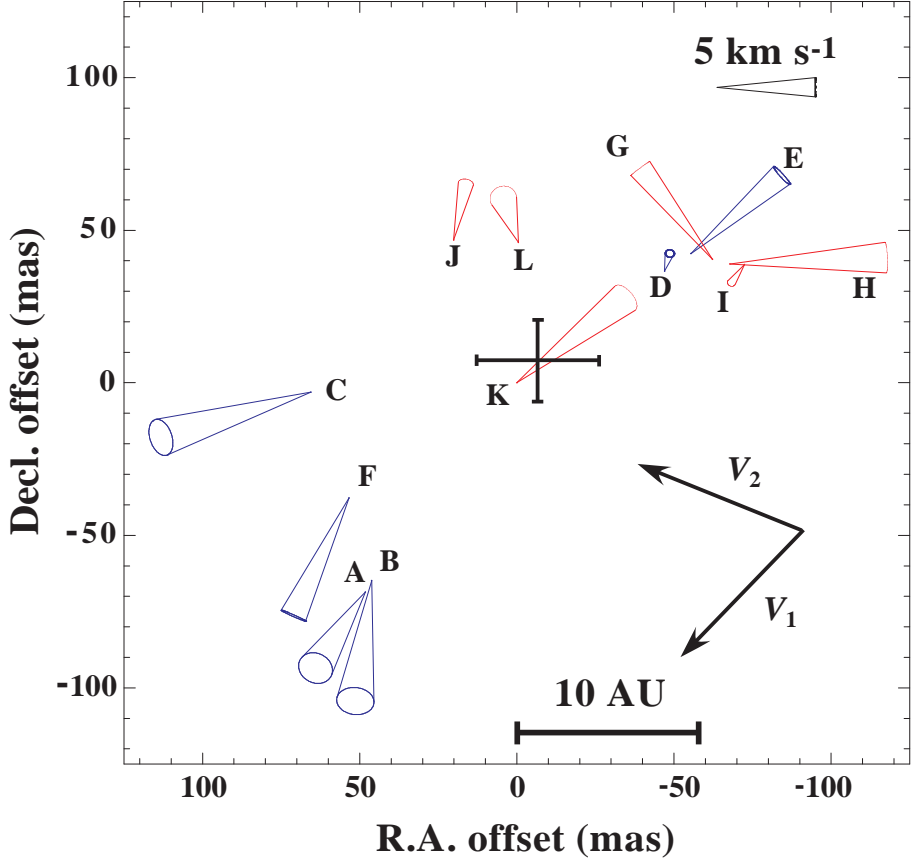}
  \end{center}
  \caption{3-D motions of maser features represented in scaled cones. The maser
           features are at the apex of the cones. The red and blue cones show 
           red- and blue-shifted features, with respect
           to the systemic velocity. The inclination of the cone indicates
           the degree of the Doppler velocity with respect to the transverse 
           velocity. The big plus-sign at the center is the model-calculated
           position of \mbox{R Crt}; the size of the plus-sign shows
           the positional uncertainties. The bold arrows are eigenvectors: 
           \mbox{\boldmath $V$}$\bf_{1}$, the largest eigenvector (the VVCM
           outflow axis) and \mbox{\boldmath $V$}$\bf_{2}$, the second-largest
           eigenvector.}
  \label{fig:Map}
\end{figure}

Figure \ref{fig:Map} shows the obtained 3-D motions of the water-maser
features.  A clear separation between the blue-shifted and red-shifted
maser features was found in the southeast and northwest directions,
which indicate bipolar outflow.  The former side of the outflow has a
velocity of $-$8.68 km s$^{-1}$ with respect to the systemic velocity,
while the latter has a velocity of 5.45 km s$^{-1}$.  Features A, B, C,
and F are within the blue-shifted side of the outflow, and spots G, H,
and I are within the red-shifted side of the suggested bipolar outflow.
Spots D and E possibly belong to the red-shifted side of the outflow,
but have blue-shifted motions.

\subsection{Objective Analysis with VVCM}

To quantify the suggested bipolarity noted from figure \ref{fig:Map}, we
used the VVCM (Variance-covariance matrix) technique, which fully and
objectively extracts the kinematic essentials, without assuming any
particular model (\cite{Bloe}).  The elements of the VVCM matrix are
expressed as

\begin{equation}
      \sigma_{ij} = \frac{1}{N{-}1}\sum_{n=1}^N(v_{i,n}{-}
      \overline{v}_{i})(v_{j,n}{-}\overline{v}_{j}) ,
      \label{eq:VVCM-EQ}
\end{equation}
where the diagonal elements are the velocity dispersions.  Here, $i$ and
$j$ denote the three orthogonal space axes (R.A., Decl., and radial
coordinates $z$), $n$ is the $n$-th maser feature of the total $N$
features, and the bar indicates the average over maser features.  The
VVCM matrix and the respective diagonalization for R Crt were obtained
as follows (in units of km$^{2}$ s$^{-2}$):

\begin{equation}
   \left(
      \begin{array}{ccc}
      18.55 & {-} 7.35 & {-} 8.10 \\
      {-} 7.35 & 15.97 & 11.43 \\
      {-} 8.10 & 11.43 & 18.79 \\
      \end{array}
   \right)
      \Rightarrow
   \left(
      \begin{array}{ccc}   
      35.82 & 0 & 0 \\
      0 & 11.62 & 0 \\
      0 & 0 & 5.86  \\
      \end{array}
   \right).
   \label{eq:VVCM-3}
\end{equation}

In figure \ref{fig:Map}, the two bold vectors show two-dimensional
projections of two unit eigenvectors with the largest eigenvalues of the
VVCM diagonalization.  The eigenvector \mbox{\boldmath $V$}$\bf_{1}$
with the largest eigenvalue, $V_{1}$= 35.82 km$^{2}$ s$^{-2}$, is the
major principal axis of the VVCM; the direction of the axis corresponds
to the outflow axis.  The axis of \mbox{\boldmath $V$}$\bf_{1}$ lies at
a P.A.  of 136$^\circ$, and is inclined at an angle of $-$39$^\circ$ with
respect to the plane of the sky with the southeast lobe directed out of
the page (towards the observer).  The second-largest eigenvector
\mbox{\boldmath $V$}$\bf_{2}$, with the second-largest eigenvalue,
$V_{2}$= 11.62 km$^{2}$ s$^{-2}$, is at P.A.  of $68^\circ$ east of
north, and is inclined at an angle of 24$^\circ$ with respect to the
plane of the sky; the vector is directed into the plane.  The
collimation factor between the two largest eigenvalues, $V_{1}$ and
$V_{2}$, is 3.08, and 1.98 between the smallest, $V_{2}$ and $V_{3}$.
On the other hand, a Monte Carlo simulation gave the following values:
$V_{1}$= 70.6 $\pm$ 52.9 at P.A.  66$^\circ$.5 $\pm$ 64$^\circ$.9, an
inclination of ${-}$4$^\circ$.5 $\pm$ 23$^\circ$.6, $V_{2}$= 24.2 $\pm$
11.5 at P.A.  34$^\circ$.2 $\pm$ 59$^\circ$.3 and an inclination of
26$^\circ$.8 $\pm$ 25$^\circ$.4.  This simulation gave the following
values of the collimation factors:  3.1 $\pm$ 1.9 for $V_{1}$, $V_{2}$
and 3.2 $\pm$ 1.8 for $V_{2}$, $V_{3}$.

We also analyzed the two-dimensional spatial variance-covariance matrix
(SVCM).  The SVCM matrix and diagonalization were obtained as follows:

\begin{equation}
   \left(
      \begin{array}{cc}
      2778.91 & {-} 1721.37 \\
      {-} 1721.37 & 1937.82 \\
      \end{array}
   \right)
      \Rightarrow
   \left(
      \begin{array}{cc}   
      4130.36 & 0 \\
      0 & 586.37 \\
      \end{array}
   \right).
   \label{eq:VVCM-2}
\end{equation}

Diagonalizing the matrix, one eigenvalue is larger than the other by a
factor of 7, indicating a slightly elongated distribution.  The axis of
elongation lies at a P.A.  of 128$^\circ$, which is close to the value
of 136$^\circ$ found for the velocity vector \mbox{\boldmath
$V$}$\bf_{1}$.  Despite the large uncertainties, the coincidence of
these two model-independent axes supports the existence of a bipolar
outflow of R Crt with a P.A.  of 136$^{\circ}$ and an inclination of
$-$39$^{\circ}$.

The smaller eigenvalues give a collimation factor of 1.98, the
inequality of the eigenvalues suggests that the outflow of R Crt is
azimuthally asymmetric around the outflow axis \mbox{\boldmath
$V$}$\bf_{1}$.  This means that transverse to \mbox{\boldmath
$V$}$\bf_{1}$ the cross section of the velocity dispersion ellipsoid
perpendicular to the outflow axis \mbox{\boldmath $V$}$\bf_{1}$ is
elongated in the direction of eigenvector \mbox{\boldmath $V$}$\bf_{2}$.
From only the 12 maser features detected, it is difficult to state the
significance of an azimuthal velocity asymmetry with an uncertainty of
98$\%$.  \mbox{\boldmath $V$}$\bf_{2}$ probably has a statistical bias
that depends on the number of maser features.  However, \citet{Bloe}
found a collimation factor of 1.54 for water masers in \mbox{W 49N}, and
suggested that magnetic fields may cause such azimuthal asymmetry.  Due
to flux freezing, ionized matter moves most easily along magnetic field
lines and experiences magnetic drag in other directions.  Therefore,
flow in a sufficiently strong field may have an elongated cross section
in velocity space, and \mbox{\boldmath $V$}$\bf_{2}$ may indicate the
direction of elongation.  \citet{Szy-a} measured magnetic fields by
observing linearly polarized OH \mbox{1667 MHz} masers of R Crt in 1995.
The results indicate that the magnetic field subtends a position angle
of $-$30$^\circ$, almost parallel to vector \mbox{\boldmath
$V$}$\bf_{1}$.  This may be a coincidence, since the magnetic field
observation lasted for $\sim$ 3 years before our observations.
Therefore, we need simultaneous observations of a magnetic field and
maser proper motions in order to confirm the correlation between the
magnetic fields and eigenvector \mbox{\boldmath $V$}$\bf_{1}$.

\subsection{Constructing the 3-D Kinematical Model}
\label{sec:CSE}

\begin{figure}
  \begin{center}
    \FigureFile(120mm,80mm){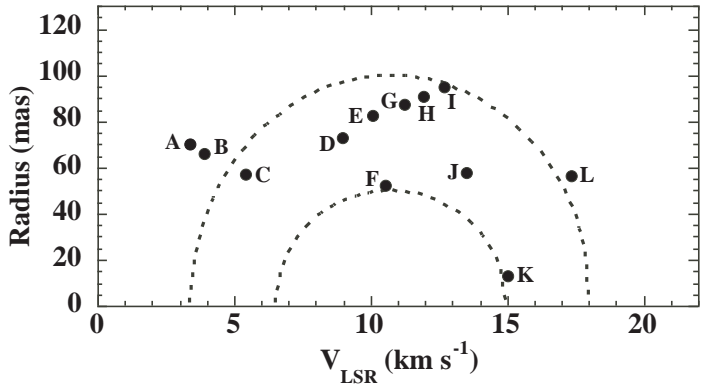}
      \end{center}
  \caption{Model fits assuming a standard expanding outflow. The radius of the
           outer edge of the water maser region is 100 mas, where the
           expansion velocity is 7.4 km s$^{-1}$. The radius of the inner 
           edge is 50 mas, where the expansion velocity is 4.3 km s$^{-1}$.}
          \label{fig:CSE}
\end{figure}

In order to derive kinematic parameters of the CSE around R Crt, the 3-D
motions of the features were fitted to an expanding outflow model.
First, the 3-D model assumes an expanding outflow where the velocity
vectors depart radially from a common center.  The details of the
fitting were described by \citet{Im-b}.  We estimated only the position
and the systemic motion of the star as free parameters, but assumed the
values of the distance to be $d$ = 170 pc and the stellar velocity to be
10.8 km s$^{-1}$.  \mbox{Table \ref{tab:RCrt-model-fit}} summarizes the
estimated parameters in the model fitting.  Second, the expansion
velocity was estimated by fitting the data to the 2-D standard expansion
model.  This model traces an ellipse on a radius-$V_{\rm{LSR}}$ plot,
and provides a reasonable description of the water maser CSE.  In the
case of the standard expanding outflow model, the projected radius, $r$,
of the maser feature at a velocity $V - V_{0}$ is given by

 \begin{equation} 
    r = 
    r_{0}\:[1 - [(V - V_{0})/V_{\rm{exp}}]^2]^{1/2}, 
    \label{eq:outflow-model} 
    \end{equation}
where $V_{0}$ is the systemic velocity of the star, $r_{0}$ the CSE
radius, and $V_{\rm{exp}}$ the expansion velocity at the CSE.  Figure
\ref{fig:CSE} shows a radius-$V_{\rm{LSR}}$ plot for R Crt and modeled
ellipses.  Least-squares fits of equation (\ref{eq:outflow-model}) using
the data (A $-$ L features) gave average values of $V_{\rm{exp}}$ = 7.4
km s$^{-1}$ and a CSE radius corresponding to $r_{0}$ = 2.6 $\times$
10$^{12}$ m.  The inner envelope fits to the parameters of $r_{i}$ = 1.3
$\times$ 10$^{12}$ m and $V_{\rm{exp}}$ = 4.3 km s$^{-1}$.  \citet{Colo}
estimated the radius of the water maser envelope of R Crt using the VLA
as 2.1 $\times$ 10$^{12}$ m with $V_{\rm{exp}}$ = 8 km s$^{-1}$ at an
assumed distance of 170 pc, which is consistent with our estimate with
compact and bright features (c.f.  \cite{Bo-d}).  The inner radius of
the water-maser envelope emission is slightly larger than the typical
radius of a dust shell with infrared emission (\cite{Danc}).  Thus,
maser features are likely to lie just above the dust shell where maser
clouds are accelerated by radiative pressure (\cite{Danc} and references
therein).  The inner radius corresponds to a few stellar radii.  The
radius of the OH emission estimated using the MERLIN array
(\cite{Szy-a}) is 3.6 $\times$ 10$^{12}$ m, where $V_{\rm{exp}}$ = 7.9
km s$^{-1}$.  \citet{Szy-a} remarked that this OH envelope should be one
of the smallest.  Material in the CSE of R Crt is accelerated to reach a
terminal velocity of $V$ = 11.0 km s$^{-1}$ (\cite{Zuck}), which was
estimated from CO observations at a maximum radius of 3.1 $\times$
10$^{14}$ m (\cite{Kaha-a}) at our assumed distance.  Thus, the
expansion velocity increases with distance.  Note, however, that each of
the different observations were performed at different epochs.


\begin{table*}

\caption{Best-fit model for the maser velocity field for R Crt.}
\label{tab:RCrt-model-fit}

\begin{center}
{\scriptsize
\begin{tabular}{lr@{}c@{$\pm$}l} \hline\hline\\[-6pt]
Parameter & \multicolumn{3}{c}{Offset} \\ \hline\\[-6pt]
Velocity: & \multicolumn{3}{c}{} \\
\hspace*{2mm}$V_{0x}$\footnotemark[*](km s$^{-1}$) ........... &5.4 & & 3.9
\\  
\hspace*{2mm}$V_{0y}$\footnotemark[*](km s$^{-1}$) ........... & $-$5.7 & &
3.9 \\
\hspace*{2mm}$V_{0z}$\footnotemark[\dag](km s$^{-1}$) ........... & 
\multicolumn{3}{c}{10.8} \\
Position: & \multicolumn{3}{c}{} \\
\hspace*{2mm}$x_{0}$\footnotemark[*] (mas) ................... & $-$9 & & 21 \\
\hspace*{2mm}$y_{0}$\footnotemark[*] (mas) ................... & 10 & & 13 \\ 
RMS residual $\sqrt{S^{2}}$ ...... & \multicolumn{3}{c}{0.47} \\ \hline\\[-6pt] 
\multicolumn{4}{c}{} \\
\end{tabular}

\footnotemark[*] Relative value with respect to the position-reference 
maser feature. \\
\footnotemark[\dag] Assuming the systemic velocity:  
$V_{0z}$ = 10.8 km s$^{-1}$. \\
}
\end{center} 
\end{table*}

The strength of the acceleration can be evaluated by comparing the
observed expansion velocity, $V_{\rm{exp}}$, with the escape velocity,
$V_{\rm{esc}}$, given by

\begin{equation} 
      V_{\rm{esc}} =
                \sqrt{\frac{2GM_{*}}{r_{0}}}, 
\label{eq:escape} 
\end{equation}
where $G$ is the gravitational constant and $M_{*}$ the mass of the
star.  Assuming a mass of R Crt of 1 $M$$_{\odot}$, at the inner radius
($r_{i}$ = 1.3 $\times$ 10$^{12}$ m, $V_{\rm{esc}}$ = 14 km s$^{-1}$,
$V_{\rm{exp}}$ = 4.3 km s$^{-1}$), $V_{\rm{exp}}$ is less than
$V_{\rm{esc}}$; consequently, masers in the inner envelope are
gravitationally bound to the star.  At the outer radius ($r_{0}$ = 2.6
$\times$ 10$^{12}$m, $V_{\rm{esc}}$ = 10 km s$^{-1}$, $V_{\rm{exp}}$ =
7.4 km s$^{-1}$), masers are marginally gravitationally bound.  Analyses
of the velocity fields of supergiants (e.g.  VY CMa, S Per, and NML Cyg)
suggest that material is, in general, bound to the star at the inner
boundary, but unbound at the outer boundary (\cite{Yates}; \cite{Rich}).
The strength of the acceleration in the case of R Crt may be less than
those of supergiants.

Finally, we discuss some relationships between the accelerations in the
maser regions, the mass-loss rates of the central stars, and the
asymmetry of the CSEs.  The acceleration is parametrized by the
logarithmic velocity gradient, $\epsilon$ = $d$(ln $v$)/$d$(ln $r$)
(\cite{Rich}).  For R Crt, within the outer limit of $r$ = 100 mas and
the inner limit of $r$ = 50 mas, the logarithmic velocity gradient is
$\epsilon$ = 0.78.  Table \ref{tab:stars-epsilon} summarizes the
velocity gradients, the mass-loss rates, and the asymmetry of the CSEs
in R Crt and in some AGB stars.  Table \ref{tab:stars-epsilon} shows
that the asymmetry of CSEs seems to occur in stars with low velocity
gradients ($\epsilon$ $<$ 0.8, ex., X Her, R Crt, NML Cyg, and VX Sgr),
but not to occur in those with high velocity gradients ($\epsilon$ $>$
1.2, ex., RT Vir and S Per).  A large number of samples would clarify
this correlation.


\begin{table*}
\begin{center}
\caption{AGB stars mass-loss rates and velocity gradients at the water-maser
regions.}
\label{tab:stars-epsilon}
\vspace{4pt}
{\scriptsize
\vspace{6pt}
\begin{tabular}{ccccc}
\hline\hline\\[-6pt] \vspace{1pt} 
Source & Type & Mass-loss rate \footnotemark[\dag] & Velocity gradient 
\footnotemark[\ddag] & Morphology \footnotemark[\S] \\
name &  & 10$^{-6}$ $M$$_{\odot}$ yr$^{-1}$  & 
$\epsilon$ &  \\
[4pt]\hline\\[-6pt]
X Her \dotfill & Semi-regular  & 0.2 \footnotemark[(1)] &
$--$ \footnotemark[(5)] & Bipolar \footnotemark[(1)] \\
R Crt \dotfill & Semi-regular  & 1.0 \footnotemark[(2)] &
0.78 \footnotemark[(6)] & Bipolar \footnotemark[(6)] \\
RT Vir \dotfill & Semi-regular  & 3.0 \footnotemark[(3)] & 
3.50 \footnotemark[(7)]\footnotemark[*] &  Spherical \footnotemark[(10)] \\
IK Tau \dotfill & Mira  & 6.0 \footnotemark[(3)] &
0.84 \footnotemark[(7)]\footnotemark[*] & Spherical \footnotemark[(11)] \\
S Per \dotfill & Supergiant  & 27.0 \footnotemark[(3)] &
1.20 \footnotemark[(7)] & Spherical \footnotemark[(12)] \\
VY CMa \dotfill  & Supergiant  & 100.0 \footnotemark[(3)] & 
0.55 \footnotemark[(7)] & Ellipsoidal \footnotemark[(13)] \\
NML Cyg \dotfill  & Supergiant  & 180.0 \footnotemark[(3)] & 
0.30 \footnotemark[(8)] & Bipolar \footnotemark[(8)] \\
VX Sgr \dotfill  & Supergiant  & 270.0 \footnotemark[(4)] &
0.50 \footnotemark[(9)] & Bipolar \footnotemark[(9)] \\
\\[-3pt]
\hline\\[-6pt] \\
\end{tabular}
}
\end{center}
{\scriptsize
\hspace*{3.0cm}
\footnotemark[*] Upper limits. \\
\hspace*{3.0cm}
\footnotemark[\dag] Mass-loss rates: \footnotemark[(1)] \citet{Kaha-b}, 
\footnotemark[(2)] \citet{Kaha-a}, \\
\hspace*{5.3cm}
\footnotemark[(3)] \citet{Yates} and references therein, \\
\hspace*{5.3cm}
\footnotemark[(4)] \citet{Eng-a}. \\
\hspace*{3.0cm}
\footnotemark[\ddag] Velocity gradients: \footnotemark[(5)] not available,
\footnotemark[(6)] this work,
\footnotemark[(7)] \citet{Yates}, \\
\hspace*{5.6cm}
\footnotemark[(8)] \citet{Rich},
\footnotemark[(9)] \citet{Chap}. \\
\hspace*{3.0cm}
\footnotemark[\S] Morphology:
\footnotemark[(10)] Imai H. private communications, \footnotemark[(11)] 
\citet{Bo-c}, \\
\hspace*{4.9cm}
\footnotemark[(12)] \citet{Rich-b}, \footnotemark[(13)] \citet{Rich-a}
}

\end{table*}

\subsection{Doppler Velocity Drifts}

One of the aims of this work is to measure the Doppler-velocity drifts
of individual water-maser features.  We found Doppler velocity drifts in
the range from $\dot{V_{z}}$ = $-$4.32 km s$^{-1}$ yr$^{-1}$ (for
feature F) to $\dot{V_{z}}$ = 0.05 km s$^{-1}$ yr$^{-1}$ (for feature
H).  Figure \ref{fig:P.M.}  and table \ref{tab:water-masers} indicate
that the blue-shifted maser features are apparently accelerated and the
red-shifted decelerated.  If the blue-shifted maser feature drifts imply
true acceleration motions, the blue-shifted side of the outflow would be
accelerated towards the terminal velocity, but the red-shifted features
would not be accelerated.  Analyses of the velocity gradients and the
escape velocities suggest that water masers are accelerated in the
envelope.  It is expected that the red-shifted masers are decelerated if
the driving source of the acceleration does not affect the red-shifted
masers at the outer limit of the water-maser envelope.

Remarkable velocity drifts were found for \mbox{features F}, the most
negative-drifted, and \mbox{feature H}, the most positive-drifted, F
being blue-shifted, and H red-shifted with respect to the systemic
velocity at $V_{\rm{LSR}}$ = 10.8 km s$^{-1}$.  Since \mbox{features F
and H} are near the systemic velocity, they move together with the star
on the plane of the sky.  \mbox{Features F and H} also have a relatively
large proper motion on the celestial plane, F having 6.8 km s$^{-1}$ and
H having 8.0 km s$^{-1}$ at the assumed distance.  On the plane of the
sky, features F and H are diagonally separated by $\sim$ 100 mas, and
the assumed stellar position lies almost at the center of the diagonal
line between the two features.  This line between features F and H
coincidentally lies at P.A.  $\sim$ 136$^\circ$, which is almost
parallel to the VVCM major axis.  It is thought that different features
do not switch on and off in a small area within the epochs of the
observations, because of the simple structure of the maser features.  We
cannot reject the possibility that the drifts do not reflect the real
motions of the circumstellar material.

\subsection{Distance to R Crt}

\begin{figure}
  \begin{center}
    \FigureFile(120mm,80mm){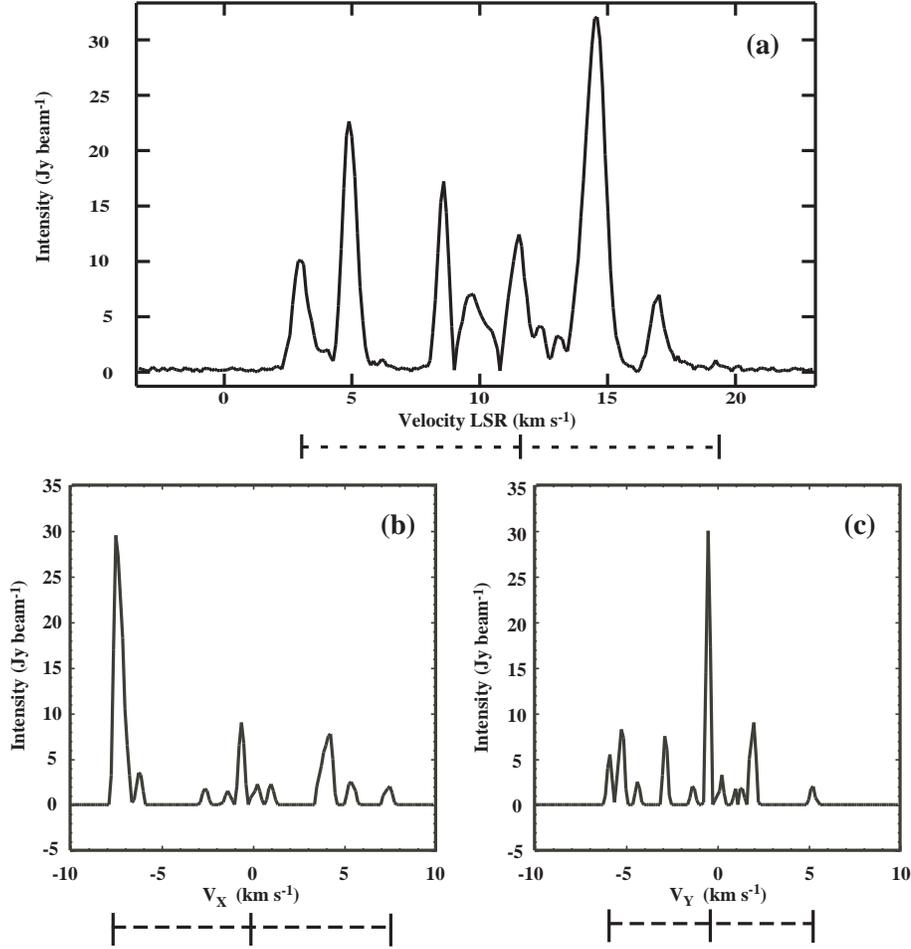}
      \end{center}
  \caption{Spectrum profiles of water masers around R Crt. (a) Total-power
           spectrum obtained at the first epoch, j98090.
          Spectrum-like profiles derived from proper-motion measurements,
          (b) on the P.A. = 136$^{\circ}$ axis  and (c) P.A. = 46$^{\circ}$
          axis directions. The dashed lines at the bottom are the velocity
          widths, and the solid vertical line at the center is the systemic 
          velocity.}\label{fig:spectrum-like}
\end{figure}

The distance to \mbox{R Crt} has been ambiguous, between \mbox{300 pc}
(\cite{Knap}) and \mbox{170 pc} (\cite{Szy-b}).  For semi-regular
variable stars, their intrinsic luminosity is unknown, and the distances
can only be crudely estimated.  Due to the small number of proper
motions, no distance measurement based on statistical parallax and a
model-fitting method was available from our data.  Instead, we made
$\lq\lq$spectrum-like profiles", flux-density profiles against proper
motions in the P.A.  = 136$^{\circ}$ axis (direction of the VVCM
eigenvector), and in the direction of the respective perpendicular axis
P.A. = 46$^{\circ}$.  In figure \ref{fig:spectrum-like}, (b) represents
the velocity profiles projected onto the directions in P.A. =
136$^{\circ}$, and (c), that projected onto the P.A. = 46$^{\circ}$
axis.

The spectrum coverages were 16.3 km s$^{-1}$ in the radial direction,
14.8 km s$^{-1}$ for P.A. = 136$ ^{\circ}$ and 10.8 km s$^{-1}$ for
P.A. = 46$^{\circ}$ axis directions at the assumed distance of
170 pc.  It is likely that the expanding flow is slightly blocked in the
northeast$-$southwest direction.  Thus, when assuming a distance of 170
pc, the spectrum-like profiles indicate that the velocity coverages in
the radial and transverse directions are roughly equal.  If a distance of
294 pc is assumed, the velocity coverages on the celestial plane would be
double, and would imply an unreliable collimation of the bipolar outflow on
the celestial plane.  Therefore, a distance of 170 pc is available for R
Crt.  Note, however, that the maser velocity coverage varies due to a time
variation in the maser features.

\section{Conclusions}

We investigated the 3-D kinematics of water masers around the
semi-regular variable star \mbox{R Crt} using data from multi-epoch
observations with \mbox{VLBI} at short intervals.  It has been verified
that most of these maser features existed for at least \mbox{80
days}.  We measured the three-dimensional velocity field, which indicates
a possible bipolar outflow.  A variance-covariance analysis applied to
the velocity field supports the bipolarity of the outflow.  The
existence of bipolar outflow suggests that a Mira variable had already
formed a bipolar outflow.  We estimated the inner and outer radii of the
water-maser region, which are consistent with those of other
observations of R Crt.  Water-maser emission in R Crt comes from regions
where dust grains are condensed, and where material within the envelope is
accelerated until reaching the terminal velocity.  It seems that within
the water-maser envelopes around R Crt, material is bound to the star in
the water-maser regions, and marginally bound outside.  We measured
the Doppler-velocity drifts of the maser features.  The blue-shifted maser
features are apparently accelerated, while the red-shifted features are
the opposite.  If the accelerations of the blue-shifted masers are real,
maser features are accelerating toward the terminal velocity.  Making
spectrum-like profiles from proper motions allowed us to verify the
distance of 170 pc to R Crt.  Similar observations are encouraged to
directly measure the accelerations in CSE to investigate envelopes around
stars and to clarify the asymmetric shapes of the CSEs of evolved stars.

\par

\vspace{1pc}\par We would like to thank an anonymous referee for
valuable comments.  We would also like to express our gratitude to each
member of the Japanese VLBI Network for their support during observations,
correlations, and data reduction.  H.I.  was financially supported by a 
Japan Society of Promotion of Science fellowship.

\end{document}